\documentclass[aps,prd,groupedaddress,preprint]{revtex4}
\usepackage{latexsym,amsfonts}
\usepackage{hyperref}
\begin{document}  

\title{The action and the physical scale of field theory}

\author{Yuri Vladimirovich Gusev}
\affiliation{Department of Physics, Simon Fraser University, 8888 University Drive, Burnaby, B.C. V5A 1S6, Canada}
\affiliation{Lebedev Research Center in Physics, Russian Academy of Sciences, Leninsky Prospekt 53, 11, Moscow 119991, Russia}

\date{\today}

\begin{abstract}
The evolution equation is used as the fundamental equation of field theory, which is described entirely by the geometry of the four-dimensional space. The evolution kernel determines the covariant effective action of physical fields by the proper time integral. This axiomatic definition introduces into dimensionless theory the universal physical scale (characteristic length). The universal scale relates the action's geometrical orders expressed in the field strength tensors. The covariant effective action is finite at any order in the curvatures and nonlocal starting from the second order. Its two lowest, local orders correspond to the cosmological constant term and the gravity action. The action of gauge fields appears in the second order term. The higher, nonlocal orders generalize the classical actions of gravity and gauge fields. The characteristic length is determined by the measured Hubble constant. The Planck scale values have no physical significance. The physical dimensionality is violated in functional integration. The dimensional regularization is an ill-defined procedure.
\end{abstract}

\maketitle

\section{Geometrical formalism for field theory}

This work began with the idea that occupied minds of physicists for about one century, 'The world is quantum'. However, this postulate inevitably brings up the fundamental problem of the emergence of classical physics and the  co-existence of the domains of quantum and classical  phenomena. This irreconcilable problem would disappear if physics could describe all observed physical phenomena - quantum and classical - by a theory derived from a unique principle. We explain here that this resolution is possible, if some mathematical and physical errors are corrected. Moreover, such a theory already exists.

Indeed, as a result of the development for about seventy years, quantum field theory has lost its features pertinent to quantum theory and become truly a geometrical theory. The field theory in this geometrical form makes it unnecessary to separate physics into classical and quantum domains because classical actions of physical fields are derived together with their generalizations which used to be considered 'quantum corrections'. This covariant field theory cannot be viewed as quantum or  classical, because it describes both 'classical' and 'quantum' physical effects, inherent to the gravity and gauge fields in their unity, with same mathematical expressions. As such it can  also be considered a unified field theory, unifying not only different physical fields, but also divided domains of physics.

The evolution of theoretical physics towards this mathematical form went for a long time until required mathematics, the evolution kernel, became available. P.A.M. Dirac summarized \cite{Dirac-PRSE1940} how the axiomatic approach to building a physical theory works. One begins with choosing a branch of mathematics, which is then developed from the first principles as far as possible before making comparison with experiments, as those maybe unavailable or erroneous. A theory developed by this method is usually associated with the constructive (axiomatic) quantum field theory, which employs the mathematical language of probability and statistics as well as the physical language of elementary particles. However, if the aim is not describing properties of individual particles as observed in scattering experiments with accelerators, but deriving the fundamental equations of physics, physical theory can be built entirely by the means of geometry and operator analysis. There are no 'quantum fields' in this form of field theory. The theory is based on the notion of geodesic distance, through the world function, and it is built only with covariant operators and tensors of physical fields. Thus, its main functional, the covariant effective action, belongs to modern geometry, which is capable of describing physical phenomena.

Physical field theory realized as the geometry of Nature poses the fundamental question: {\em Where does the physical scale come from, if the field theory is dimensionless?} There is only one way to gain the length - the only physical quantity in geometry - the axiomatic principle of  the covariant action of physical (observable) fields. This is the way for dimensionless (expressed in pure numbers) mathematical functions to became physical quantities measured in physical units, i.e. to gain physical dimensionality. Traditionally, field theory was made dimensional by the introduction of the notion of {\em particles} that possess energy and mass and move in space and time. But, this mechanism inevitably retracts field theory to the {\em mechanistic} form that  has long inhibited the progress of physics because the notion of a particle cannot be even defined. Indeed, a physical particle is an object with a measured (effective) size, thus, it is rather a physical body that occupies a spacetime domain than a 'point-like' particle. Developing the mechanics of particles as bodies that obey certain (relativistic and quantum) rules has not led to a consistent physical theory. Therefore, it is reasonable to reject this theoretical paradigm and return  to phenomenological field theories built with physical observables. 

The  technical subject of the present paper is  the geometrical formalism for field theory. This formalism is derived with help of the evolution equation on the Riemannian spin manifold with the gravity and gauge field connections \cite{DeWitt-book1965,DeWitt-book2003,BarVilk-PR1985}. Its main functional, the covariant effective action, is expressed in terms of geometrical characteristics of the base manifold and tensors of physical fields. In the  obsolete form it is known as the Schwinger-DeWitt expansion \cite{DeWitt-book1965}. The covariant perturbation theory (CPT), used for the derivation of the covariant nonlocal effective action in its general form, was proposed in \cite{GAV-Gospel,GAV-CERN1992,GAV-CQG1992}, developed in the series of papers \cite{CPT1,CPT2,CPT3,CPT4} and later used to build the general theory of radiation \cite{GAV-lectures2007,GAV-AP1998}. This {\em mathematical} theory is fully covariant and expressed in terms of phenomenological fields linked to physical observables.

Neither the Planck constant, nor the Planck length can serve as the scaling parameters in field theory. The 'bare' action of quantum field theory has physical nature entirely different from the 'one-loop covariant effective action', therefore, the two cannot be summed up to one mathematical expression. Consequently, we are forced to abandon the view on the effective action as a 'quantum correction'. The mathematical form and the physical meaning of the covariant nonlocal effective action make it the {\em universal action} of physical fields. It becomes obvious that the effective action can only be introduced by the {\em axiomatic} definition. The classical actions of gravity and gauge fields are just a small part of the universal effective action, whose other parts are important for physics as well. We will refer sometimes to the covariant  effective action  simply as the action.

In Appendix \ref{NewSI} we analyze M. Planck's works and the New SI  of physical units (2019) and conclude that the Planck's values do not play any role in physics because they are arbitrary. Therefore, the Planck length, Planck's energy, etc cannot be used as the needed physical scale. In Appendix \ref{hbar} we show that the Planck constant cannot appear in field theory due to its physical dimensionality. For this reason, functional integration (in contrast to the 'path integral') is not valid, and the 'loop expansion' of the effective action does not exist. In Appendix \ref{dimregerr} we explain why the technique of dimensional regularization is inconsistent. However, phenomenological physical theories in contrast to quantum field theories cannot contain infinite quantities (divergences), thus, the procedures of regularization and renormalization are irrelevant to them.

Then we revisit the covariant perturbation theory and express the hypothesis that the covariant 'one-loop effective action', when it is defined axiomatically, presents the phenomenological action of physical fields. From this definition, the unique length scale emerges and appears throughout the action. Two lowest in the curvatures orders  of the evolution kernel ('the heat kernel') are also present.  Physically, they represent the Hilbert-Einstein gravity action and the cosmological constant term. Their appearance from  the effective action method was proposed fifty years ago by traditional techniques. However, the result was divergent and therefore indefinite. With the correction, which eliminates erroneous divergences, the action gets a different physical meaning and will find many applications in experimental physics and technology. 

\section{Evolution kernel in the covariant perturbation theory} \label{heatkernel}                                       

We begin with standard notations and definitions \cite{BarVilk-PR1985,CPT1,CPT2}. The spacetime (or rather space) has the Euclidean metric signature  and the integer dimension $D$. All field theories that describe physical phenomena  possess the Laplace type differential operator \cite{BarVilk-PR1985},
\begin{equation}
     \hat{F}(\nabla)= \Box \hat{1} +  \hat{P} - \frac{1}{6}R\hat{1}.
     \label{operator}  
\end{equation}
The Laplace-Beltrami operator (Laplacian) is constructed of the covariant derivatives,
\begin{equation}
 \Box \equiv g^{\mu\nu} \nabla_{\mu} \nabla_{\nu},  \label{Laplacian}
\end{equation}
whose indexes are contracted by the matrix, $g^{\mu\nu}$, the inverse of which
is the metric of the Riemannian manifold (for the details and conditions, see \cite{BarVilk-PR1985}). 
The Ricci scalar in (\ref{operator}) was not originally present in \cite{CPT1}, it was introduced in \cite{CPT2} only for the conformal scalar field model, which is non-physical.
The potential term $\hat{P}$ in (\ref{operator}) is an  local function of fields, its behaviour is specified \cite{CPT2}. The 'hat' notation stands for the indexes of internal  degrees of freedom, e.g. of non-Abelian gauge fields. Therefore, the Utiyama-Yang-Mills fields \cite{Utiyama-PR1956} are included in the expressions below.

Let us stress the fact, that it is absolutely {\em unnecessary} to introduce the notion of dimensionless (quantum) fields on which this operator would act. All operators act on dimensional (physical) fields expressed as field strength tensors, as obvious in \cite{CPT1,CPT2}. A quantity, which not does not appear in final results and is  not  used even in derivations, is redundant. This operator formalism can be considered as a modern, covariant version of the operator analysis invented in the 19th century by Oliver Heaviside \cite{Heaviside-PRSL1892,Heaviside-PRSL1893}. In the spirit of Heaviside's physical mathematics, the Laplacian (\ref{Laplacian}) is used in the covariant perturbation theory as a variable, whose properties are well specified. 

The covariant derivatives in (\ref{Laplacian}) contain both metric and gauge field connections, which are characterized by the commutator curvature,
\begin{equation}
(\nabla_{\mu} \nabla_{\nu}-
\nabla_{\nu} \nabla_{\mu} )  =
\hat{\mathcal{R}}_{\mu\nu}, \label{commutator}
\end{equation} 
whose spin-tensor properties are embedded into the matrix notation and specify an explicit form of the tensor $\hat{\mathcal{R}}_{\mu\nu}$. For example, for the Abelian gauge field it is proportional to the Maxwell electromagnetic tensor \cite{GAV-AP1998,GAV-lectures2007,Shore-NPB2002}. The commutator curvature $\hat{\mathcal{R}}_{\mu\nu}$ is one type of the basic field strength tensor, while the Ricci tensor $R_{\mu\nu}$ and the local potential $\hat{P}$ are two others. The full set of these curvatures is denoted by,
\begin{equation}
\Re \equiv (R^{\mu\nu},\hspace{3mm} \hat{\cal R}_{\mu\nu}, \hspace{3mm} \hat{P}). \label{Re}
\end{equation}

We seek the kernel, $\hat{K} (s| x,x')$, of the {\em evolution} equation (erroneously called the 'heat equation') \cite{DeWitt-book1965,DeWitt-book2003,CPT2},
\begin{equation}
	\frac{\partial}{\partial s}
	\hat{K} (s| x,x')
	=\hat{F}(\nabla^x)
	\hat{K} (s| x,x'),         \label{heateq}
\end{equation}   
with the initial condition,
\begin{equation}
	\hat{K} (s| x,x')
	=\hat{\delta}(x,x'), \ {\sigma(x,x')/s} \gg 1, \label{delta}
\end{equation} 
where $s$ is a parameter with the dimensionality $m^2$ called the {\em proper time}. The Ruse-Synge {\em world function} \cite{Ruse-PLMS1931,Synge-book1960} is the fundamental physical quantity defined by the equation,
\begin{equation}
	2 \sigma(x,x')=g^{\mu\nu} (x) \nabla^{(x)}_{\mu}\sigma(x,x') \nabla^{(x)}_{\mu}\sigma(x,x'). \label{sigma}
\end{equation} 
In the Cartesian coordinates, $\sigma(x,x')$ is half a square of the geodesic distance between spacetime points $x$ and $x'$ \cite{Synge-book1960}, thus, the physical theory is built with a physical observable, the {\em distance}. The meaning of the initial condition (\ref{delta}) of the evolution equation (\ref{heateq}) is different from the one accepted in \cite{CPT2,Gusev-NPB2009}. The proper time is a dimensional parameter, therefore, the declared asymptotics, $s \rightarrow 0$, cannot be  defined mathematically. The proper time must be compared with a scale similar in physical dimensionality, this can only be the geodesic length defined by the world function $\sigma(x,x')$.  

The {\em fundamental} solution of  the evolution equation (\ref{heateq}) with the initial condition (\ref{delta}) is the zeroth order of the evolution kernel  \cite{DeWitt-book2003,CPT2}, which can be viewed as the covariant delta function in the {$D$-dimensional} spacetime,
\begin{equation}
\hat{K}_0 (s|x, x')  =
\frac{g^{1/4}(x) g^{1/4}(x')}{(4 \pi s)^{D/2}} 
\exp\left(-\frac{\sigma(x,x')}{2s }\right) \hat{a}_0 (x,x'), \label{K0}
\end{equation}
where $\hat{a}_0(x, x')$ is the propagator of parallel transport \cite{DeWitt-book1965,BarVilk-PR1985,CPT2}.  The function (\ref{K0}) does not exist for $s=0$, and the asymptotics $[{\sigma(x,x')}/s] \gg 0$ does not imply the proper time values goes to zero. In the kernel of the evolution  equation $\hat{K} (s|x, x')$, the positive proper time $s$ is the parameter external to the spacetime variables. 

Let us stress that the evolution equation (\ref{heateq}) is principally different from the equations of diffusion and heat propagation \cite{Vladimirov-book1971}. Due to its significance for physics, the evolution equation should be considered the fundamental equation of theoretical physics. By displacing the Schr\"odinger equation, which is somewhat similar in a form, but  different in physical meaning, the evolution equation introduced fifty years ago a new paradigm of physical mathematics. Instead of seeking solutions as wave functions (or quantum fields) and then using them to derive some probabilistic properties of physical observables, the evolution kernel directly delivers a mathematical solution which is expressed in terms of physical fields.

In order to obtain the action, we only need to know the functional trace of the heat kernel,
\begin{equation}
{\textrm{Tr}} K (s)  =
\int {\mathrm d}^{D}  x \,  {\textrm{tr}}\,  \hat{K} (s|x,x), \label{TrK}
\end{equation}
which, beside the matrix trace over the field indexes, {\em tr}, assumes the coincidence of spacetime points and the integral over the whole spacetime domain, $\mathbb{R}^D$. The metric's determinant, $g^{1/2}(x)$, is included in $K(s|x,x)$.  The functional trace of the kernel, ${\textrm{Tr}} K (s)$, is a dimensionless functional, in contrast to the  kernel's coincidence limit, $K(s|x,x)$, whose depends on the spacetime dimension, $D$. In this paper we refer to (\ref{TrK}) as the evolution kernel.

The algorithms of the covariant perturbation theory  \cite{CPT2} were used for computing the evolution kernel.  The CPT is the covariant expansion of (\ref{TrK}) in asymptotically flat spacetime, in orders of the curvatures that are expressed as nonlocal tensor invariants \cite{CPT2,BGVZ-JMP1994-bas}. The CPT evolution kernel contains an infinite number of covariant derivatives acting on the curvatures, thus, it is a nonlocal expression. The zero and first orders of the evolution kernel are local expressions \cite{CPT2}. Starting from the second order, the trace of the heat kernel is nonlocal. Its form, up to the third order in the curvatures (for $D \leq 5$) is \cite{BGVZ-JMP1994-bas,CPT4},
\begin{eqnarray}
\mathrm{Tr}\, K(s)&=&
\frac{1}{(4 \pi s)^{D/2}} \int \! {\mathrm d}^{D} x\, g^{1/2}(x)
{\rm tr} \left\{ \hat{1} + s \hat{P}
+ s^2 \sum_{i=1}^{5} {f}_{i}(s, \Box_1, \Box_2) \Re_1 \Re_2(i) \right.
\nonumber\\&&\left. \mbox{}
	+ s^3 \sum_{i=1}^{29} F_{i}(s, \Box_1, \Box_2 , \Box_3 ) 
	\Re_1 \Re_2 \Re_3(i)
	+ {\rm O}[\Re^4] \right\}.  \label{TrK3}
\end{eqnarray}
The five structures quadratic in the curvatures, derived in \cite{CPT2}, are reviewed below. The third order in the curvatures contains 29 cubic structures that were computed and studied in Refs.~\cite{CPT4,BGVZ-JMP1994-bas,BGVZ-JMP1994-asymp}.  The CPT calculations are carried out with the accuracy ${\rm O}[\Re^n]$, so that the highest computed order contains terms of up to the {$(n-1)$-order} in the curvatures {\em explicitly}. Therefore, the validity of the covariant perturbation theory obeys the condition, 
\begin{equation}
\nabla \nabla {\Re} \gg \Re^2.
\end{equation}
Only the denominator of the kernel's prefactor (\ref{TrK3}) depends explicitly on the spacetime dimension, which greatly simplifies the field theory calculations.  

The explicit expressions for the form factors of (\ref{TrK3}) can be found in \cite{CPT2,CPT4}. The form factors, $f_i$, as well as the third order ones, $F_i$, are {\em analytic} functions of the dimensionless operator-valued variables, 
\begin{equation}
\xi_i=-s\Box_i. \label{xi}
\end{equation}
They act on the nonlocal tensors invariants, and  the index of the Laplacian in (\ref{xi}) indicates the local curvature it is acting on, i.e.,
$ \Box_2 R_1 \hat{P}_2\equiv R(x) (\Box \hat{P} (x))$.  The expressions resulting from these operations are taken at the coincident spacetime point, $x$, which is a variable of the spacetime integral (\ref{TrK3}). The kernel with the coincident points, $\hat{K}(s|x,x)$, up to the second order in the curvatures, is similar in the structure to (\ref{TrK3}) and can be derived from it  \cite{Gusev-NPB2009}. The second order form factors in (\ref{TrK3}) are expressed via the basic form factor, 
\begin{equation} 
	f(-s\Box) \equiv \int_0^1\!\! {\mathrm{d}}\alpha\:  
	\exp(-{\alpha(1-\alpha)(-s\Box)}). \label{ff} 
\end{equation}  
Two form factors with subtractions are given by the integrals,
\begin{equation}
\frac{f(-s\Box)  - 1}{(s\Box)} =
\int_{0}^{1} \mathrm{d}\alpha_1  
\alpha_1 (1-\alpha_1)
\int_{0}^{1} \mathrm{d}\alpha_2  
\exp \Big( -\alpha_2 \alpha_1 (1-\alpha_1)(-s\Box) \Big), \label{ffsub1}
\end{equation}
\begin{equation}
\frac{f(-s\Box)  -  1 -  \frac16 (-s\Box)}{(-s\Box)^2} =
\int_{0}^{1} \mathrm{d}\alpha_1  
\alpha_1^2 (1-\alpha_1)^2
\int_{0}^{1} \mathrm{d}\alpha_2  
\alpha_2
\int_{0}^{1} \mathrm{d}\alpha_3
\exp \Big( - \alpha_3 \alpha_2 \alpha_1 (1-\alpha_1) (-s\Box) \Big). \label{ffsub2}
\end{equation}

Let us now use the method of \cite{Mirzabekian-PLB1996} and replace the tensor basis for $\mathrm{Tr} K(s)$ of \cite{CPT2} based on the Ricci tensor by the basis defined via the Weyl tensor, $C_{\alpha\mu\beta\nu}$ (from this point all expressions are in {\em four} dimensions). Namely, a new tensor quantity was defined,
\begin{equation}
C_{\mu\nu} \equiv \frac{2}{\Box} \nabla^{\beta} \nabla^{\alpha} C_{\alpha\mu\beta\nu}, \label{Weyl}
\end{equation}
which is related to the Ricci tensor, $R_{\mu\nu}$, via the perturbative expression \cite{Mirzabekian-PLB1996},
\begin{equation}
R_{\mu\nu} = C_{\mu\nu} + \frac{1}{3} \nabla_{\mu} \nabla_{\nu}  \frac{1}{\Box} R + \frac{1}{6} g_{\mu\nu} R + \mathrm{O}[\Re^2]. \label{Ricci}
\end{equation}
Thus, the new basis differs from the old one \cite{CPT2} only by the first term,
\begin{eqnarray}
\Re_1\Re_2({1})&=&C_{1\,\mu\nu} C_2^{\mu\nu}\hat{1}, \label{RR1}
\\[\baselineskip]
\Re_1\Re_2({2})&=&R_1 R_2\hat{1},  \label{RR2}
\\[\baselineskip]
\Re_1\Re_2({3})&=&\hat{P}_1 R_2,   \label{RR3}
\\[\baselineskip]
\Re_1\Re_2({4})&=&\hat{P}_1\hat{P}_2,   \label{RR4}
\\[\baselineskip]
\Re_1\Re_2({5})&=&\hat{\cal R}_{1\mu\nu}\hat{\cal R}_2^{\mu\nu}. \label{RR5}
\end{eqnarray}
In the $C_{\mu\nu}$ tensor basis, only form factors of the pure gravity terms change, 
\begin{eqnarray} 
\tilde{f}_1(\xi) &=& \frac{1}{2} f_1 (\xi),\\
\tilde{f}_2(\xi) &= & f_2(\xi) + \frac{1}{3} f_1 (\xi),
\end{eqnarray}
where $f_i(\xi)$ are the original form factors of \cite{CPT2}.
The new set of form factors becomes,  
\begin{eqnarray} 
	\tilde{f_1}  (-s\Box) &=& 
    \frac{f(-s\Box)  -  1-  \frac16   s\Box}{(s\Box)^2},    \label{ff1}\\ 
	\tilde{f_2}  (-s\Box) &=& \frac{1}{24}   \left[ 
  	\frac{1}{12}\,  f( -s \Box )  -  
     \frac{  f( -s \Box ) - 1}{ s \Box } +
  	 5 \frac{  f( -s \Box ) - 1 - \frac16 s \Box }{ (s \Box)^2} 
				\right],                                    \label{ff2}\\ 
	\tilde{f_3} ( -s \Box ) &=& f_3 =
  \frac1{12}\,   f( -s \Box ) - \frac12\,   \frac{f( -s \Box )-1}{ s \Box }, 
  \label{ff3} \\     	
	\tilde{f_4}   ( -s \Box ) &=& f_4 = \frac12  \,  f( -s \Box ), 
 \label{ff4} \\
	\tilde{f_5}   ( -s \Box ) &=& f_5 = \frac12  \, \frac{f( -s \Box )-1}{ s \Box }. 
 \label{ff5}
\end{eqnarray}  
The short and large proper time asymptotics of the second order form factors can be found in \cite{CPT2}.  The new basis leads to the simplifications of the third order form factors in the effective action considered in \cite{Mirzabekian-PLB1996,BMZ-grqc1995}.

\section{Covariant action: local and nonlocal terms} \label{effectiveaction}                                       

The effective action was introduced to quantum field theory by J.S. Schwinger \cite{Schwinger-PR1951}, who also first proposed to use the Euclidean spacetime \cite{Schwinger-PR1959}. The {\em covariant} effective action was invented by B.S. DeWitt in order to apply this method to gravity and gauge field theories \cite{DeWitt-book1965}.  It is expressed in terms of phenomenological fields (called the 'expectation value' fields in \cite{GAV-lectures2007}) and serves as a generating functional of the field theory currents, including the energy-momentum tensor  \cite{GAV-AP1998} that enters the effective equations.  

The geometrical formalism of the effective action (its erroneous was the Schwinger-DeWitt technique) is entirely different from the Feynman's path integral. It is a differential technique \cite{Schwinger-talk1973} in contrast to the integral technique of R.P. Feynman. Nonetheless, the effective action is often assumed to be derived from the functional integral \cite{DeWitt-book2003}, but this is an erroneous statement as explained in Appendix \ref{hbar}. We propose that the action be defined axiomatically. This is it, the 'one-loop effective' action is {\em not} the first order term in a series in the powers of the Planck constant. Therefore, it is {\em not} 'a quantum correction' to some classical action, rather it contains a classical action in itself. In fact, the quantity called 'action' in the field theory \cite{DeWitt-book2003} is {\em not} the action by its physical meaning and physical dimensionality, therefore, the Planck constant that appears in quantum {\em mechanics} does not appear in field theory. This error entered theoretical physics when mathematicians generalized the Feynman's path integral by the means of functional analysis, apparently, first in \cite{Gelfand-UMN1956}.

Thus, {\em axiomatically} the covariant action is the following proper time integral of the evolution kernel,
\begin{equation}
-W (l^2) \equiv
 \int_{l^2}^{\infty}\! \frac{{\mathrm d} s}{s}\,  
	\mathrm{Tr} K(s).         \label{effac}
\end{equation} 
Because the evolution kernel (\ref{K0}) does not exist for $s=0$ there must be a positive lower limit of the proper time integral, denoted as $l^2$ in (\ref{effac}), with the dimensionality $m^2$. This limit should not be be 'small', because there is no other quantity in this integral to compare with. Therefore,  $l^2$ is arbitrary finite. The functional $W$ is dimensionless, because both the evolution kernel, $\mathrm{Tr}\, {K}(s)$,  and the proper time integration measure in (\ref{effac}) are dimensionless. 

The solution for the evolution kernel (\ref{TrK3}) is substituted into (\ref{effac}). After the proper time integration, the sum in the curvatures of the evolution kernel (\ref{TrK3}) turns into the corresponding sum of the covariant action, which depends on the  $l^2$-scale,
\begin{equation}
-W(l^2)=
\sum_{n=0}^{\infty}
(l^2)^{(n-2)} 
W_{(n)}(l^2).  \label{Wmu2}
\end{equation}
The action in four dimensions, including the third order in the curvatures, was computed in \cite{CPT2,BGVZ-NPB1995,CPT4}. That expression changes now by the addition of two  local terms that  were previously eliminated by the dimensional regularization \cite{BarVilk-PR1985,CPT2} (see also Appendix \ref{dimregerr}),
\begin{eqnarray}
-W(l^2)
&=& \frac{1}{(4\pi)^2}\int\! \mathrm{d}x^4\, g^{1/2}(x) \, \mathrm{tr}\,
		\Big\{
	l^{-4} \, \frac{1}{2} \hat{1} + l^{-2}\, \hat{P}
\nonumber\\&& \mbox{}
	+ l^0\, \sum^{5}_{i=1}\gamma_i(-\Box_2 l^2)\Re_1\Re_2({i}) +
	\nonumber\\&& \mbox{}
	+  l^2\, \sum_{i=1}^{29} G_{i}(-\Box_l^2 l^2, -\Box_2 l^2 , -\Box_3 l^2 ) 
	\Re_1 \Re_2 \Re_3(i)
	+ {\rm O}[\Re^4] \Big\}.  \label{effac3}
\end{eqnarray}
The third order form factors of the action (\ref{effac3}) are now dimensionless (as different from the representations of \cite{CPT4}) and contain the $l^2$ factor. The third order in the curvatures gains therefore the overall factor $l^2$. The functional $W$ is defined up to a multiplier (the calibration constant), which should be determined by experiment. The value of the scaling constant $l^2$ is not defined by the theory, however, the relations between different orders of the covariant action, which are already established by the existing physical laws and constants, could give the value of $l^2$.

Let us now revisit the computation of the second order form factors of (\ref{effac3}). In this order of the action, the  integral (\ref{effac}) of the basic form factor (\ref{ff}) has the form,
\begin{equation} 
	\gamma(-\Box l^2) \equiv  \int_{l^2}^{\infty}\! \frac{{\mathrm d} s}{s}\, 
	\int_{0}^{1} \mathrm{d}\alpha \exp \Big( - \alpha (1-\alpha)s(-\Box) \Big). \label{eaff} 
\end{equation}
After introducing the new dimensionless parameter, $t$,
\begin{equation} 
	t = s \alpha (1-\alpha) (-\Box)
\end{equation}
and changing the order of integrals, (\ref{eaff}) can be expressed  as, 
\begin{equation} 
	\gamma(-\Box l^2) =
	\int_{0}^{1} \mathrm{d}\alpha_1 
	 \int_{z}^{\infty}\! \frac{{\mathrm d} t}{t}\, 
	\exp ( - t ), \label{effacg1} 
\end{equation}
where the exponential integral has the lower limit, 
\begin{equation}
z=\alpha (1-\alpha) (-\Box l^2). \label{z}
\end{equation}
The exponential integral  can be solved by standard mathematical tools, e.g.  \cite{Olver-book1974},
\begin{equation}
	\int_{z}^{\infty}\! \frac{{\mathrm d} t}{t}\, 
	\exp ( - t ) =
	- \mathcal{C} - \ln(z) 
	- \sum_{n=1}^{\infty} \frac{(-1)^n z^n}{n \cdot n!},  \label{Eilimit}
\end{equation}
where the Euler constant is $\mathcal{C} \approx 0.577216$.
Substituting this  back to (\ref{effacg1}) and doing the $\alpha$-integral gives the following expression,
\begin{equation}
	\gamma(-\Box l^2) = - \ln \big(-\Box l^2 \big) - 
	\mathcal{C} + 2 + \frac{1}{6}(-{\Box}{l^2}) 
	+ \mathrm{O}\big[ -\Box l^2  \big].         \label{ffgea1}
\end{equation}
In (\ref{ffgea1}) we keep the first term of the sum in (\ref{Eilimit}), because it adds up to the {\em third} order of the covariant action (\ref{effac3}) and so on. In this truncated form, the expression (\ref{ffgea1}) is valid as the asymptotic,
\begin{equation}
\big(-\Box l^2 \big) \ll 1, \label{boxlimit}
\end{equation}
while the expression (\ref{Eilimit}) holds for {\em arbitrary}  $\big(-\Box l^2 \big)$.  

We use the integral representations (\ref{ffsub1})-(\ref{ffsub2}) to compute two other integrals for the basic form factors with subtractions,
\begin{eqnarray}
\eta \big(-\Box l^2 \big)&= &
\int_{l^2}^{\infty}\! \frac{{\mathrm d} s}{s}\,
\frac{f(-s\Box)  - 1}{s\Box)}, \label{effacg2}\\
\theta \big(-\Box l^2 \big) & = & \int_{l^2}^{\infty}\! \frac{{\mathrm d} s}{s}\,
\frac{f(-s\Box)  -  1 -  \frac16 (-s\Box)}{(-s\Box)^2}. \label{effacg3}
\end{eqnarray}
Their asymptotic expansions are given by,
\begin{eqnarray}
\eta \big(-\Box l^2 \big) & =& 
- \frac{1}{6}\Big[ \ln \big(-\Box l^2 \big) + 
	 \mathcal{C}\Big] + \frac{4}{9} - \frac{1}{60} \Box l^2 
	+ \mathrm{O}\big[ (-\Box  l^2)^2 \big],         \label{ffgea2}\\
\theta \big(-\Box l^2 \big) & =& 
 - \frac{1}{60}\Big[ \ln \big(-\Box l^2 \big) + 
 \mathcal{C} \Big] + \frac{23}{450} - \frac{1}{840} \Box l^2 
	+ \mathrm{O}\big[ (-\Box  l^2)^2 \big].         \label{ffgea3}
\end{eqnarray}
The full list of the second order form factors in the action (\ref{effac3}) is,
\begin{eqnarray} 
	\gamma_1 (-\Box l^2) &=& - \frac{1}{60}\Big[ \ln \big(-\Box l^2 \big) + 
	 \mathcal{C} \Big] + \frac{23}{450} - \frac{1}{840}\Box l^2,\label{eag1}\\ 
	\gamma_2 (-\Box l^2) &=& -\frac{1}{1080} - \frac{1}{7560} \Box l^2,\label{eag2}\\ 
	\gamma_3 (-\Box l^2) &=& -\frac{1}{18} - \frac{1}{180}\Box l^2, \label{eag3}\\     	\gamma_4 (-\Box l^2) &=& - \frac{1}{2} \Big[ \ln \big(-\Box l^2 \big) + 
	 \mathcal{C}\Big] + 1 - \frac{1}{12}\Box l^2,\label{eag4}\\
	\gamma_5 (-\Box l^2) &=& - \frac{1}{12}\Big[ \ln \big(-\Box l^2 \big) + 
	 \mathcal{C}\Big] + \frac{2}{9} - \frac{1}{120}\Box l^2.\label{eag5}
\end{eqnarray}  
The second order of the action (\ref{effac3}), given by the  expressions (\ref{eag1})-(\ref{eag5}) together with the nonlocal tensor invariants (\ref{RR1})-(\ref{RR5}), reproduces the result of \cite{Mirzabekian-PLB1996}. The difference is that the expressions above are only the asymptotics $(-\Box l^2) \ll 1$ of the exact form factor (\ref{effacg1})-(\ref{Eilimit}). Let us emphasise that neither  ($l^2 \rightarrow 0$), nor  ($\Box \rightarrow 0$) are correct forms of this asymptotics, and the asymptotics (\ref{boxlimit}) was really meant under the notation ($\Box \rightarrow 0$) in the covariant perturbation theory works.  The opposite limit, $(-\Box l^2) \gg 1$, might also be worth studying.

In the new tensor basis (\ref{RR1})-(\ref{RR5}), the second order terms with the scalar Ricci tensor in the action (\ref{effac3}) are local as seen from (\ref{eag2})-(\ref{eag3}) (the local term  $\hat{P} R$  is known since Ref.~\cite{CPT2}). The properties of the covariant action related to the scalar Ricci tensor are studied  in \cite{Mirzabekian-PLB1996}.

The second order of the covariant effective action (\ref{effac})  contains the Utiyama-Yang-Mills field action (\ref{RR5}) with the form factor (\ref{eag5}). Its appearance indicates the existence of observable non-Abelian fields which is confirmed by the collider experiments and by the lattice numerical simulations. In fact, lattice simulations are closer to the correct description of subnuclear physics than traditional analytical tools of quantum chromodynamics, because the characteristic length scale introduced by the lattice cell size corrects the consequences of infinities erroneously introduced in quantum field theory: 'ultraviolet divergences' of the momentum space formalism stem from the second order of the covariant effective action which was presumed to be divergent \cite{CPT2}.

\section{The physical scale and the gravity action}

The content of the preceding section belongs to geometrical  methods of the operator analysis.  Let us now connect these mathematical expressions with physics starting with the lowest orders in the curvature.   The first local term of the covariant action (\ref{effac3}) is trivially universal.  The second local term is determined by the potential, $\hat{P}$.  Therefore, this order is identically zero, when $\mathrm{tr}\hat{P}=0$, the condition which is true only for conformal scalar field models.  The operator (\ref{operator}) for these models contains the Ricci scalar term with the coefficient (-1/6), which was the reason for singling it out. Thus, the covariant action for conformal scalar field models possesses no first order term. A field theory that does not generate the gravity action (as explained below) is clearly non-physical.

In the currently accepted theory of the interactions of elementary {\em particles}, quantum chromodynamics, fundamental fields are massless spinors \cite{PDG-PRD2018}. These fields correspond to quarks that do  not exist as free (unbound) particles. It is quite natural then not to use the notion of spinor fields, but to assign mathematical properties of spinors to spacetime itself. The mathematical formalism of spin manifolds was developed by R. Penrose and W. Rindler \cite{Penrose-book1984-1}, but a large body of literature exists. Our methods requires only the knowledge of the commutator curvature (\ref{commutator}). Indeed,  in this formalism derivatives always act on on spacetime and gauge field  tensors \cite{CPT2}.

The square of the covariant Dirac operator equals the Laplace-Beltrami operator and the Ricci scalar with the numerical coefficient (-1/4). This operator was introduced first by E. Schr\"odinger \cite{Schroedinger-PADW1932,Friedrich-book2000}, but often associated with the names of A. Lichnerowicz and B.S. DeWitt \cite{DeWitt-book1965}. To obtain this form  for the operator $\hat{F}(\nabla)$ by the algorithmic rules \cite{BarVilk-PR1985}, the potential term, $\hat{P}$ (which also includes the gauge field tensor \cite{GAV-lectures2007,Shore-NPB2002} eliminated in the first order by the trace operation), should give the contribution,
\begin{equation}
  \mathrm{tr} \hat{P} =  - \frac{1}{12} R \, \mathrm{tr}  \hat{1}. \label{Pterm}
\end{equation}

With the substitution of (\ref{Pterm}) to (\ref{effac3}), two local terms of the covariant action gain opposite signs. The Hilbert-Einstein action of the gravity theory possesses the  Ricci scalar curvature term without any dimensional factor, e.g. \cite{Landau-fieldV2}. This scaling can be achieved by multiplying the whole action by the factor of $l^2$. Since according to  the main postulate of the proposed theory, the covariant action is known only up to a factor determined by experiment, this operation does not change physics. Thus, the lowest orders of the action, normalized by other unessential factors as, 
\begin{equation}
	\bar{W} (l^2) \equiv {(4 \pi)^2} 12 l^2 \, W (l^2).
\end{equation} 
have the final form, 
\begin{equation}
	\bar{W} (l^2) =   \int\! \mathrm{d}x^4\, g^{1/2}
 	\mathrm{tr}\hat{1} \, \Big\{6/l^2   - R  + l^2 {\rm O}[\Re^2]
	 \Big\}.  \label{gravity}
\end{equation}
It is natural to  identify the zero order term of $\bar{W} (l^2)$ with the cosmological constant, while the first order term represents the Hilbert-Einstein action. They emerge in the covariant action, which previously was deemed existing only from the second order \cite{CPT2}.

It is obvious that the gravity action cannot be added (in models for the study of black hole or early Universe physics) to the effective action without a dimensional factor, because they have different dimensionalities: the action is dimensionless, $m^0$, while the dimensionality of the gravity action is $m^2$ (this fact adds a dimensional factor to the gravity action in Ref.~\cite{Mirzabekyan-JETP1994}). The Planck length is usually employed as the needed dimensional parameter, however, we argue  that the Planck length has an arbitrary value (see Appendix \ref{NewSI}). The physical scale $l^2$ emerged from the mathematical solution above, but its physical consequences are yet to be studied.  Absolute numerical coefficients of these two terms are not defined, because the  value of the scale parameter $l^2$ can only be measured, but not derived.  The local orders as well as all pure gravity terms of the higher orders of the action (\ref{effac}) have the same relative coefficients for any spin group, because the trace of the unity matrix, $\mathrm{tr}\hat{1}$, factorizes out of these terms (\ref{gravity}). 

A large number of {\em ad hoc} extensions of general relativity have been proposed  and some of them could be tested with astrophysical observations \cite{Berti-CQG2015}. The most viable candidates for modified gravity represent the Hilbert-Einstein action with the addition of higher orders in curvatures and/or nonlocal terms. Some of them are postulated while others are justified by various field models, but all such modifications contain free parameters. The generalization proposed here (\ref{effac3}) is derived from the unique axiomatic principle, therefore delivers numerical coefficients of the local and nonlocal terms in the higher orders in the curvatures. This result may lead to the development of an {\em axiomatic} cosmological theory based on the field theory. The currently dominant cosmological {\em model}, whose observational foundation remains uncertain \cite{Herouni-RNASA2007}, is facing many unresolved problems and should be eventually replaced by a proper theory.

The expression (\ref{gravity}) lets us obtain a value of the universal physical scale from the measured cosmological constant $\Lambda$,
\begin{equation}
 6/l^2 = \Lambda. \label{CC}
\end{equation}
The value of the Hubble Constant recently measured by the Supernova team with the Hubble Space Telescope is $H_0 =  73.48 \pm 1.66\ (km/s)/Mpc$ \cite{Riess-AJ2018}. Then, a length corresponding to this   observed quantity is the Hubble length, $c/H_0 \approx 1.26 \cdot 10^{26}\ m$. This number can be used as a proxy for the size of the Universe. The cosmological constant can be calculated according to the currently accepted model \cite{PDG-PRD2018} as $\Lambda \approx 1.88 \cdot 10^{-52}\ m^{-2}$. If no cosmological model is assumed, this number loses the factor of 3. Consequently, by Eq.~(\ref{CC}) the length constant is,
\begin{equation}
l \approx 1.8 \cdot 10^{26}\ m, \label{scale}
\end{equation}
which is close to the Hubble radius, but in this dimensional analysis only the order of magnitude matters. The relation between the cosmological constant and the size of the Universe was conjectured first by Dirac \cite{Dirac-book2005}. It seems quite natural that the  physical scale of field theory is supplied by the fundamental distance of the Nature, which makes the theory of {\em observed} physical phenomena closed. Assigning a fixed value to the parameter $\mu^2=1/l^2$ in \cite{Mirzabekyan-JETP1994} does not change the computation of the form factors. 

These physical conjectures naturally follow from the mathematical derivations, however, they should be further studied before attempting to develop physical theories based on them.

\section{Summary} \label{summary}

In the present paper, dimensional analysis is applied to the covariant perturbation theory in order to correct its solution and re-consider its physical meaning. Let us summarize the main points.  
\begin{itemize}
\item
The evolution equation is the fundamental equation of field theory which delivers its kernel determined by geometrical properties of the four-dimensional Riemannian spin manifold.
\item
The covariant effective action, axiomatically determined the evolution kernel, is the main functional of physical fields expressed in a covariant form.
\item
The proper time integral which defines the covariant effective action introduces the physical scale into dimensionless field theory.
\item
This scale (characteristic length) is present in all orders of the action, which is in finite and nonlocal.
\item
Two lowest orders of the covariant action are local represent the cosmological constant term and the Hilbert-Einstein action of gravitation.
\item
The higher orders in the curvatures of the covariant effective action are generalization of the actions of gravitational and gauge fields.
\item
The value of the characteristic length is determined by the measured Hubble constant.
\item
The Planck values (Planck length, Planck mass, etc) have no physical significance.
\item
Planck constant cannot appear in field theory for dimensional reasons which invalidates the functional integration.
\item
Dimensional regularization technique is mathematically and physically inconsistent.
\end{itemize}

\section{Discussion} \label{discussion}

{\em The covariant perturbation theory is a non-perturbative method.}

The covariant perturbation theory is the only tool to obtain the nonlocal kernel of the evolution equation in a general form. Nevertheless, its name caused some confusion resulting in misunderstanding of its results and wrong applications of them. As a matter of fact, CPT is {\em not} a perturbation theory even though its derivation begins with the  expansion in the operator's  perturbation \cite{CPT1}. However, the procedure, which makes it covariant, uses  nonlocal nonperturbative substitutions \cite{CPT2}, thus, the  final result is not only covariant, but also fully nonpertubative (in the traditional language of field theory). This is it, every geometrical order of the CPT action contains the gravity and gauge field connections (including implicitly  the 'coupling constants')  to the infinite order. As a result, the covariant nonlocal effective action should be considered as {\em a sum} of the nonlocal tensor invariants of the increasing order, each of which is responsible for its own domain of physical phenomena, separated by the universal physical scale.

{\em The proper time is the key variable of the evolution equation.}

The proper time method was discovered and developed in parallel in physics and mathematics during the 20th century. The early work in physics was done by V. Fock \cite{Fock-pt1937}, who recognised that the proper time is a new variable in relativistic physical theory. The proper time method became popular after the Schwinger's work \cite{Schwinger-PR1951}, but it also appeared in the work of  Y. Nambu \cite{Nambu-PTP1950}. The evolution equation (\ref{heateq}) used to be called the 'heat equation', but this name creates confusion with the Fourier equation of heat conductivity. We suggest to use the term 'evolution kernel' for the trace of the kernel of the evolution equation (\ref{TrK3}). The evolution equation is widely studied in mathematics literature, where it is used to develop the theory of geometrical flows. Some of its applications in geometry via the Ricci flows were studied by R. Hamilton \cite{Hamilton-JDG1993} (cf. Ref.~\cite{BGVZ-JMP1994-bas}).The recent developments of the Ricci flows theory helped prove the Poincare Conjecture \cite{Tao-arxiv2006}. The evolution kernel under the name of the 'heat kernel' has been extensively studied in mathematical physics. This subject was advanced by I. Avramidi \cite{Avramidi-book2015}, where it is also applied to solve long-standing problems in financial mathematics. 

{\em The world function is a fundamental concept of physical theory.}

The conclusion that the covariant action based on the evolution kernel is the fundamental mathematical object of physical theory is supported by its very structure. Indeed, the functional trace (\ref{TrK}) of the kernel $\hat{K}(s| x, x')$ is the only combination similar to the traditional Lagrangian. Furthermore, the evolution kernel is built with the world function (\ref{sigma}), i.e. with the {\em geodesic distance of spacetime}. This fact conforms with the omnipresent idea that physical theories must be {\em geometrical}. The proper time fraction in the effective action definition (\ref{effac}) makes it dimensionless, in agreement with the proposed postulate that physical functionals must be {\em dimensionless}. The presence of the large characteristic length in the covariant action creates the hierarchy of physical scales that separates domains of physical phenomena, thus, allowing  building specialized  theories limited in the scope of description. 

{\em The length scale must be present in the effective action.}

Contrary to the generally accepted opinion, the 'classical' action in quantum field theory does not  have the dimensionality of the action (as it does in quantum {\em mechanics}), in fact, it is $meter^2$ (see Appendix \ref{hbar}). The dimensionality of the Planck constant is $Joule \cdot second$, but nor time, neither energy appear yet in the  {\em geometry} of Euclidean spacetime. The Planck length is not relevant to physics at all (see Appendix \ref{NewSI}). Therefore, neither constant can appear in the calculations above. However, through the action definition (\ref{effac}) we gain another physical scale with the correct dimensionality of $m^2$. This fact does not change most of the mathematical derivations previously done in the covariant perturbation theory.

{\em The universal physical scale can only be measured.}

The only thing that could be said about the physical scale is its measured value (\ref{scale}). Its possible relation to the cosmological constant is discussed above, but this discussion relies on the validity of the presently accepted cosmological model and should be re-analyzed, when this model were to change. In dimensionless field theory we have yet to gain the notion of mass (or energy), therefore, no comparison with other physical constants can be made yet.  The length scale, $l^2$, which enters the form factors of the nonlocal covariant action as $1/\mu^2=l^2$, could specify the assumed 'massless' approximation \cite{GAV-lectures2007,GAV-AP1998} and apparently does not change its mathematical derivations.

{\em The effective action in the Euclidean space is  transferred to the Minkowski spacetime.}

The covariant action is computed in spacetime with the Euclidean metric signature, i.e. in the {\em four-dimensional space}. The transfer to the Minkowski spacetime is easy, because the local terms of (\ref{effac3}) are insensitive to the metric signature. The procedure to transform the nonlocal terms was derived in \cite{CPT1} and it consists of 1) doing the variation over the metric and 2) replacing all Euclidean Green functions in the obtained energy-momentum tensor with the retarded ones.

{\em The effective action contains the classical actions of gravity and gauge fields.}

The effective action, obtained by solving the evolution equation with the Laplacian-based operator, is a phenomenological functional of the gauge field and gravity strength tensors. Its different orders in the curvatures (different geometrical orders) are related (scaled) to each other by the unique length scale, which naturally  appears in the theory. The 'classical' actions (the Hilbert-Einstein action of gravity theory and the Maxwell-Heaviside action of electrodynamics) are correspondingly the first and the second order terms of the effective action. The higher order contributions, that were previously deemed to be 'quantum' actions, add up in (\ref{effac3}) with the scaling factor, which is another fundamental (defining) physical constant. This proves the main conjecture of this work, that the covariant nonlocal effective action is the universal action of all physical fields.

Three years after the present paper was completed in June 2016, we found the work that treats  the cosmological constant in a somewhat similar way. Anderson and Finkelstein showed \cite{Finkelstein-AJP1971} that the cosmological constant must be present in the action of classical gravity for mathematical reasons. This statement agrees with our derivations although mathematics justifying this term is different. They further suggested that the presence of the cosmological constant is equivalent to the existence of the characteristic ('fundamental atomic') length in physics. However, without the gauge field connections of Riemannian manifold the conjecture of \cite{Finkelstein-AJP1971} could not be mathematically implemented.

{\em On the Sakharov-DeWitt proposal of induced gravity.}

The first order term of the covariant action (\ref{effac}) is  the action of the general relativity theory of gravitation. The emergence of the gravity action in quantum field theory was first suggested by A.D. Sakharov \cite{Sakharov-DAN1967} in the  proposal later known as the {\em induced gravity}. Even though Sakharov's work used mathematics similar to ours (the 'heat kernel' from mathematical literature), it could not be completed without resorting to the language of field theory. In field theory, the dimensional scale present in the action is fundamentally physical. It is not an auxiliary parameter ('regulator') that should be somehow eliminated, but must be enter physical quantities. The emergence of the cosmological constant and the gravity action from the 'heat kernel' was also suggested in the 1963 Les Houches School lectures by B.S. DeWitt \cite{DeWitt-book1965}. The misconceptions of divergences and renormalization have survived to the present time \cite{DeWitt-book2003}, nevertheless, we suggest to call this derivation of the gravity action the {\em Sakharov-DeWitt mechanism}. 

{\em The  generalized electrodynamics emerges within the effective action.}

The second order of the effective action (\ref{effac3}), with the form factor (\ref{RR5}) and the commutator curvature (\ref{Re}), for the Abelian gauge fields contains electrodynamics modified by nonlocal and higher order contributions. This modification is neither surprising, nor new. The generation of the action of electrodynamics within quantum field theory was proposed first by Ya.B. Zeldovich \cite{Zeldovich-JETPL1967}, who  applied Sakharov's idea of induced gravity to gauge fields. However, the completion of this idea was not possible at that time, because one has to abandon the concepts of quantum vacuum and particles to achieve the mathematically acceptable solution. Modifications of the  action of classical electrodynamics have been developed  in the past, e.g. see the nonlocal electrodynamics of F. Bopp \cite{Bopp-AdP1943}, but this subject, its history and relation to Ref.~\cite{GAV-AP1998} lies beyond the scope of the present paper.

{\em The operator analysis was discovered by Heaviside.}

The history of the form factors of the nonlocal action can be traced as far back as to the 19th century, when the operator analysis was invented and used by O. Heaviside \cite{Heaviside-PRSL1892,Heaviside-PRSL1893}. Rejected at first by mathematicians as not rigorous and even erroneous,  Heaviside's physical mathematics was later confirmed by other mathematical methods and eventually incorporated into mathematics \cite{Erdelyi-book1962}. The operator calculus as a computational tool of quantum field theory that was developed in the later 20th century can rightly be considered as the modern reincarnation of Heaviside's methods.

{\em There are no infinite quantities in phenomenological theories.}

Above we showed that non-physical infinities appeared in the effective action because of a trivial mathematical error. Indeed, if one introduces the zero proper time limit in (\ref{effac}), which contradicts mathematics because it is equivalent to the division by zero in the evolution kernel (\ref{TrK3}), this error appears as the infinite term, see Appendix~\ref{dimregerr}. This and other infinite terms are then discarded by a postulate that no power like divergences are present in the effective action. Thereby, two lower order terms, the gravity theory action, are absent. Even though some other terms in the basic form factor (\ref{ffgea1}) appear correctly, the rest of this series, which contributes to higher order $l^2$ terms, cannot be recovered. The crucial physical meaning of the universal length scale (\ref{CC}) is lost as well.

The same infinities appear as 'ultraviolet divergences' in quantum field theory formulated in the phase-space, where they are ridden off by the procedures of regularization and renormalization. However, even Feynman wrote about renormalization  \cite{Feynman-book1985}, p. 128: {\em ''I suspect that renormalization is not mathematically legitimate''.}  Dirac was more emotional about the theory of quantum electrodynamics  which he initialized \cite{Dirac-book1978}, p. 36: {\em ''Sensible mathematics involves neglecting a quantity when it turns out to be small - not  neglecting it just because it is infinitely great and you do not want it!''} Creator of modern electrodynamics and operator analysis Heaviside  was harshly critical about 'senseless' mathematics \cite{Heaviside-PRSL1893}, p. 121: {\em ''So all solutions of physical problems must be in finite terms or in convergent series. Otherwise nonsense is made.''} Forty years ago Schwinger began publishing his three volume book ''Particles, sources, and fields''  that presented the 'source theory', {\em ''to which the concept of renormalization is foreign''}  \cite{Schwinger-book1989}, p. ix, because it is a phenomenological theory of elementary particles without divergences. In general, phenomenological  theories deal with observable physical quantities, which are always finite. 

{\em The Hawking radiation problem stimulated the development of new mathematics.}

The development of the covariant perturbation theory \cite{CPT1,CPT2,CPT3,CPT4} began in late 1980s with the aim of creating a mathematical theory for tackling the hypothetical effects of particle creation  by the electromagnetic ('Schwinger effect' \cite{Schwinger-PR1951}) and gravitational ('Hawking radiation' \cite{Hawking-CMP1975}) fields. The latter one prompted this development because the gravitational collapse is an evolution problem. The solution of this problem requires taking into account the back-reaction of emitted radiation on the metric. Hawking wrote about the particle creation by black holes \cite{Hawking-CMP1975}, p. 216: {\em ''Because it  is a non-local process, it is probably not reasonable to expect to be able to form a local energy-momentum tensor to describe the back-reaction of the particle creation on the metric.''} The needed {\em nonlocal} energy-momentum tensor was obtained in \cite{GAV-AP1998} from the covariant effective action of \cite{CPT4}. 

However, the problem of Hawking radiation cannot be solved rigorously even with this new mathematics because the black hole metric is still assumed as an initial condition, while it should really be derived as a solution. Furthermore, experimental studies show that the  laws of thermal radiation cannot adequately describe observed quantities, i.e. these laws are neither universal, nor exact, e.g. \cite{Gusev-RJMP2014}. Therefore, thermal radiation physics cannot provide  principles for building fundamental theories. Yet, the motivation to resolve the theoretical paradox of black holes radiation led to the creation of new mathematics and the attainment of the physical action that can solve real problems of experimental physics. In science the  investigation of an encountered paradox often leads to the development of new physics.

{\em Physics is an experimental science.}

The principal difference of the proposed meaning of the effective covariant action from the one assigned to it in quantum field theory must be  emphasized. The {\em phenomenological} physical theory is always built with variables that are physical {\em observables}. Such a theory delivers phenomenological quantities, i.e. it describes {\em and} predicts physical phenomena. The discovery of the universal action of physical fields in form of the 'one-loop effective action' was made more than two decades ago, \cite{GAV-lectures2007}, p. 759:  {\em ''The effective action does not refer even to quantum field theory. It is an action for the observable field, and its implications may be valid irrespective of the underlying fundamental theory''}. However, the existence and the type of a hypothetical fundamental theory cannot be experimentally deduced and verified, because its notions, quantum field and vacuum state, are not observable in principle. Three centuries ago I. Newton wrote about such notions in his {\em hypotheses non fingo} phrase \cite{Newton-Scholium1729}: {\em ''But hitherto I have not been able to discover the cause of those properties of gravity from phænomena, and I frame no hypotheses. For whatever is not deduc’d from the phænomena, is to be called an hypothesis; and hypotheses, whether metaphysical or physical, whether of occult qualities or mechanical, have no place in experimental philosophy''}.

\section*{Acknowledgments} 
I am grateful to Max Planck Institute for Gravitational Physics (Albert Einstein Institute) at Potsdam-Golm, Germany for support and hospitality during several visits.

\section{Appendix A. Planck's values and the New SI of physical units}  \label{NewSI}

In this paper we use the term 'physical dimensionality' to avoid confusion with the spacetime dimension. Dimensionality appears from experiments that measure a physical quantity by comparing it with a reference quantity - a physical unit (etalon). The operation of measurement brings the scale to physics. Physical theories unify natural phenomena by  physical laws that relate different physical quantities to each other in mathematical expressions. As a result, the number of independent physical quantities is rather small. The International Committee of Weights and Measures ({\em Bureau international des poids et mesures, BIPM}) agreed that  the minimal number of physical quantities (units) needed for all practical purposes is seven. Until 20018 BIPM recommended to use these SI (Syst\`eme International) physical units in physical and engineering sciences.

However, scientific community will switch in May 2019 to use {\em the New SI} of physical units \cite{Mills-PTRSA2011,Stock-Metro2019}. After this revolutionary change, the New SI has physical constants with the {\em exact} (fixed) numerical values, while physical units will be measured with some uncertainty. The General Conference on Weights and Measures (CGPM) on November 16, 2018 adopted the resolution \cite{Stock-Metro2019} to use the units of kilogram, ampere, kelvin with experimental uncertainties. These physical units are now defined by exact values of the Planck constant, the elementary charge, the Boltzmann constant \cite{Stock-Metro2019}. The fundamental physical constants are called now the {\em defining} constants to reflect their physical meaning. The newly adopted system is similar to the system of physical units (called the natural or Planck's system) proposed by Max Planck in  \cite{Planck-AdP1900}, Sect. 26, but different from it by a specific choice of physical constants caused by  the progress of physics over the century. Let us briefly review the Planck's system to explain some misunderstanding in the literature.

The  system of natural physical units is also presented in M. Planck's book ``The theory of heat radiation'' \cite{Planck-book1914}, Sect. 164 ``Natural units''. Planck explains that physical units used to be chosen {\em ad hoc} and based on material artefacts that are special or relevant to the existing intelligent life, in the given conditions. However, {\em ``with  the aid of the two constants $h$ and $k$ which appear in the universal  law of radiation, we have the means of establishing units of length,  mass, time, and temperature, which are independent of special  bodies or substances, which necessarily retain their significance  for all times and for all environments,  terrestrial and human or  otherwise, and which may, therefore, be described as ``natural  units''.} In other words, with two new constants, called later the Planck constant and the Boltzmann constant, the system of physical units became {\em coherent}, in the language of modern metrology, because the number of (fundamental) units is equal to the number of (fundamental) constants. Then, a particular choice of absolute numerical values is not relevant to physics, only relations  among them, expressed as physics laws, are.

Planck suggested to use this arbitrariness by selecting values of the fundamental physical constants in a predetermined way. The simplest choice is to assign values 1 to all four  constants. As a side effect, physical units of this new system would have unusual numerical values, {\em if expressed by the SI units}. Planck used the Newton's constant of gravitation, $G_{\mathsf{N}}$, as one of fundamental constants. The use of $G_{\mathsf{N}}$ can partially explain the popularity of the Planck's system  in the gravity and high energy theory literature.  In the New SI, the Newton's constant is a derivative constant. Besides, it is a physical constant measured with low precision, whose value is plagued by discrepancies between results of different experiments \cite{Quinn-PTRSA2014}. 

In the Planck's system, the physical unit of length would be \cite{Planck-book1914}, $l_{\mathrm{Planck}}=\sqrt{G_{\mathsf{N}}/(h c^3)}$, and other units expressed similarly through the SI constants.
The values of such units are commonly referred to as the Planck' values, e.g. the Planck length is the value of the unit of length in the Planck system expressed by the SI metre,
$ 1\ \mathrm{unit\ of\ length}\ \mathrm{(Planck\ system)} \approx 3.99 \cdot 10^{-35} \ \mathrm{meter}\ \mathrm{(SI)}$.
However, the only reason to assign ones to fundamental physical constants instead of any other numbers is that 1 is the smallest number in any numeral system, which is clearly not a physical reason. Selecting any other numbers would arbitrarily change all Planck's values: they have no fundamental significance for physics.

M. Planck writes further, {\em ''These quantities retain their natural {\rm meaning} as long as  the law of gravitation and that of the propagation of light in  a vacuum and the two principles of thermodynamics remain  valid; they therefore must be found always the same, when  measured by the most  widely  differing intelligences according to the most widely differing methods''} \cite{Planck-book1914}. Under the words {\em `` terrestrial and human or  otherwise''} above, emphasized by the phrase {\em `` the most  widely  differing intelligences''} in this paragraph, Planck apparently meant that any human society or, perhaps, extraterrestrial intelligent life that developed a system of physical units in the simplest possible way, i.e. by assigning ones to the fundamental physical constants, would find the same physical etalons used by any other intelligence, when they compare the implementations of their physical units (like the  metal bar of one metre that used to be an etalon of SI).  

However, the adopted New SI is  similar to the Planck's system by the principal idea, but implemented {\em differently}. In the New SI, physical units form the hierarchy of inter-dependencies. The first defining constant is the unperturbed ground state hyperfine transition frequency of the caesium 133 atom ${{}^{133}Cs}$ $9,192,631,770\ Hz$. It defines the unit of second, which does {\em not} depend on any other unit. With help of the second defining constant, the speed of light in vacuum, the unit of length is defined, and so on. In other word, natural numbers, the most fundamental mathematics, form the very foundation of modern physics. Of course, other value of other physical frequency could be used for the constant defining the unit of time, and the value of the speed of light is also arbitrary, and so on. Assigning unity to the frequency constant would give a peculiar value of $1.09 \cdot 10^{-10}\ s$ to the New SI unit of time, and so on. It is obvious, that any such choice is arbitrary and obtained numbers are irrelevant to physics.

The previously used SI was set up according to historical conventions. It could certainly be inconvenient to start using unusual (very small or very large) units. Therefore, historical conventions are approximately kept in the New SI \cite{Stock-Metro2019}.  For example, in the New SI the metre is  not exact, but very close to the SI metre. The principal difference of the New SI from the SI is that the values of physical units are known only {\em approximately}, while the physical constants remain exact, the opposite agreement used to hold in the SI.

Summarizing, in his works Max Planck did not declare that the numerical values of new physical units in the proposed natural system expressed via the physical units of the traditional system would have special physical meanings. The Planck length, mass, etc are just a curiosity: changing the scale of the system of physical units does not change physics.  Assigning any physical significance to the Planck's values contradicts the core idea of the Planck' system, the recently adopted New SI  and the scale-free nature of field theory.

\section{Appendix B. No Planck constant in field theory} \label{hbar}

The Planck constant plays a special role in quantum mechanics, but it is also deemed to be significant in quantum field theory. Let us analyze this confusion with the dimensional analysis and show that the Planck constant cannot even appear in field theory. First of all, the Planck constant is the {\em defining} constant of the physical unit of mass as explained in Appendix \ref{NewSI}. Its value is fixed permanently \cite{Stock-Metro2019} as, 
\begin{equation}
	\hbar \equiv 1.054\ 571\ 800 \times 10^{-34}\ kg \cdot m^2 \cdot s^{-1}.
\end{equation} 
The former experimental uncertainty of $\hbar$ is  assigned now to the unit of mass, which can be measured with an increasing precision.

It is customary in the quantum field theory literature and related areas to omit in equations the symbols of fundamental (defining) physical constants. The reason usually given for such {\em textual} simplification is the choice of the Planck's (or natural) system of physical units \cite{Planck-book1914}, p. 174 (discussed in Appendix \ref{NewSI}). However, this shortening of mathematical notations is not really related to the Planck's system. Indeed, the Planck constant is given value 1 in this system, however, its physical dimensionality ($J \cdot s$)  remains. Thus, we could as well agree on not writing explicitly the symbol $\hbar$ (or other physical constants) in {\em any} other system of physical units. After calculations are done, One must restore all defining physical constants  and check the consistency of the  physical dimensionality of derived mathematical expressions. The  values of physical constants are  needed only for experimental verifications of theory's predictions. 

The dimensionality of the action in {\em classical mechanics} is the product of energy and time. The dimensionality of the action in {\em quantum mechanics} is by definition equal to the dimensionality of the  action in  classical mechanics. The Lagrangian formalism for quantum mechanics was proposed by Dirac \cite{Dirac-PZS1933}. This spacetime approach to quantum theory was later developed by Feynman \cite{Feynman-RMP1948} as the formalism of {\em path integral} (the integral over all allowed trajectories of a massive particle) \cite{Zinn-Justin-Scholar2009}. It was built using the classical mechanics action for a 'quantum' particle, $S_{\mathsf{QM}}$, \cite{Feynman-RMP1948}: {\em ''The contribution from a single path is postulated to be an exponential whose (imaginary) phase is the classical action (in units of $\hbar$) for the path in question''}. Therefore, the Planck constant renders the quantum mechanical action dimensionless.  Consequently, the argument of the path integral's exponential function, 
$
	\exp \Big( \frac{i}{\hbar} {S_{\mathsf{QM}}} \Big),
$
is dimensionless as required.

During the development of quantum field theory, the path integral was generalized to the {\em functional integral}, e.g. \cite{Gelfand-UMN1956}, and the Planck constant in its exponent was taken for granted. However, when physical constants are omitted in derivations, the  dimensionality of physical expressions is easily confused. This is it, the ('bare') action of quantum field theory, $S_{\mathsf{QFT}}$, is built in the way similar to the quantum mechanics one, 
\begin{equation}
	S_{\mathsf{QFT}}= -1/2\int d^4 x \, g^{\mu\nu}\, 	
	(\nabla_{\mu} \varphi) (\nabla_{\nu} \varphi).
\label{QFTaction} 
\end{equation}
However, the quantum field $\varphi$ is dimensionless, and the theory is {\em relativistic}, which means the time coordinate is always multiplied by the speed of light constant $c$, i.e. all four dimensions of {\em spacetime} are measured in meters. Therefore, the action (\ref{QFTaction}) has the dimensionality of squared meters. This means that the Planck constant, whose dimensionality is $kg \cdot m^2 \cdot s^{-1}$, cannot be used to make $S_{\mathsf{QFT}}$ dimensionless. Thus, it cannot appear in the exponent of functional integral like it is commonly done, e.g. \cite{Zinn-Justin-Scholar2009} (last section).  This discrepancy makes the field of functional integration invalid. An error of this type can occur when mathematicians try to formalize a physical theory. Mathematics works with pure numbers, i.e. dimensionless variables and constants, thus, special care is required in {\em mathematization} of physics.

If one formally expands any physical quantity in the powers of $\hbar$, then the coefficients at each order of this expansion would have different dimensionalities. The terms of this expansion  are functions (or functionals), and in order for this expansion to be convergent, these functions must be small, compared to the corresponding power of the Planck constant, for {\em  any} values of this quantity's variables. It is clear that this is not true in general. The theory of quantities deprived of physical dimensionality is mathematics. For a series expansion to be universally valid, a  dimensionless function must be expanded in a {\em dimensionless variable}. 

The Schwinger-DeWitt (geometrical) formalism of the efective action is entirely different from the Feynman's path integral, nevertheless, the effective action is often assumed to be derived by functional integration \cite{DeWitt-book2003}. However, functional integration is invalid, and the $\hbar$-expansion (called the 'loop expansion') of the effective action \cite{DeWitt-book1965,DeWitt-book2003} cannot even exist. The effective action (\ref{effac}) is axiomatically defined by \ref{effac}, and the term 'one-loop' is dropped off its name.

\section{Appendix C. Dimensional regularization is erroneous}  \label{dimregerr}

When infinite quantities (called 'ultraviolet divergences') were encountered in quantum field theory, the method of regularization was used to explore them by introducing an auxiliary parameter  \cite{DeWitt-book2003}. In the 'effective action' method, the {\em dimensional} regularization used to be employed \cite{BarVilk-PR1985,CPT2}. However, in the covariant perturbation theory divergences are an artefact of improper computations and should not appear at all. This  fact is sufficient for physics, nevertheless, we show that dimensional regularization is erroneous and gives wrong answers.

The origin of the dimensional regularization is phase space integrals  of the Hamiltonian formalism of the elementary particle theory. The Feynman's path integral formalism uses spacetime coordinates and particle momenta as variables. The phase space integrals explicitly depend on the spacetime dimension $D$. After making the spacetime dimension an arbitrary (complex) parameter and computing  the {$D$-dimensional} integrals, physical spacetime dimension $D=4$ is restored \cite{Leibbrandt-RMP1975}. 

In the covariant perturbation theory, which uses mathematics of the evolution kernel, variables are the world function's covariant derivatives \cite{Synge-book1960,BarVilk-PR1985,GAV-lectures2007}, and the action is a functional of the nonlocal tensor invariants \cite{CPT2,GAV-lectures2007}. The second order form factors are expressed by the exponential integral (\ref{effacg1}). The proper time integral is computed by standard mathematics \cite{Olver-book1974}, with the spacetime integral defined in {\em four} dimensions. As we have seen above, the resulting covariant action (\ref{effac3}) is finite and nonlocal. 

However, the dimensional regularization is not only unnecessary, it is also theoretically inconsistent.  Let us reconsider the derivation of the basic form factor at $D=4$ \cite{CPT2}. Its definition in the covariant perturbation theory \cite{DeWitt-book2003,BarVilk-PR1985,CPT2} (and the related literature), the lower limit of the proper time integral (\ref{effac}) is $s=0$. Then, the form factor in the spacetime with $D$ dimensions admits the solutions \cite{CPT2},
\begin{equation}
\gamma(-\Box) \equiv 
 \int_{0}^{\infty}\! \frac{{\mathrm d} s}{s}\, 
	f (-s \Box) =
 (-\Box )^{D/2-2}
 \frac{\Gamma(2-D/2) \Gamma(D/2-1)^2}{\Gamma(D -2)}, \label{dimreg}
\end{equation} 
where $f (-s \Box)$ is defined by Eq.~(\ref{ff}), and $\Gamma$ is the gamma function. The effective action form factor in this form (\ref{dimreg}) has the dimensionality $m^{4-D}$, which is compensated with the dimensionality of the spacetime integration measure $\mathrm{d}x^D$. They make, together with the tensor structures (\ref{RR1}), a {\em dimensionless} functional.  The  expansion of (\ref{dimreg}) in the parameter, $\epsilon \equiv 2- D/2\ll 1$, which is small in the four-dimensional asymptotic, is then taken in dimensional regularization. The following $\epsilon$-expansion is then utilized, 
\begin{equation}
	(-\Box  )^{D/2-2}\equiv 
	(-\Box )^{-\epsilon}=
	\exp \big( -\epsilon \ln(-\Box ) \big) = 
	1 - \epsilon \ln (-\Box) + {\mathrm{O}}[\epsilon^2], \ \epsilon \ll 1. \label{Boxe}
\end{equation}
However, this expression is erroneous: its l.h.s. has dimensionality $m^{D-4}$, while every terms of its r.h.s. is dimensionless. Of course, the logarithmic function must have a dimensionless  argument, but even when one substitutes $(-\Box)$ with $(-\Box l^2)$ by hand, as is done in the literature, the expression (\ref{Boxe}) is still wrong. It is obvious that the expansion (\ref{Boxe}) makes sense (converges) only if its terms are convergent. However, due to the logarithmic function this expansion is valid only in the limit, $(-\Box l^2) = \mathrm{O} [1]$, i.e., at the single point, not for arbitrary values of the argument, as claimed and used. This fact rejects the usual argument that the physically important logarithmic behaviour of the form factor is  still correctly derived by dimensional regularization, it is not.

The product of (\ref{Boxe}) with the expansions of gamma functions in (\ref{dimreg}) gave the final result \cite{CPT2},
\begin{equation}
	\gamma (-\Box) =  \frac{1}{\epsilon} - 
	\ln (-\Box) -  \mathcal{C} 
	+ 2 + {\mathrm{O}}[\epsilon], \ \epsilon \ll 1,
	\label{erreaff}
\end{equation}
which is also erroneous. This expression differs from (\ref{ffgea1}) by the additional divergent term, $1/\epsilon$. This term of (\ref{erreaff}) appears from the {\em division by zero} infinity in the proper time integral, i.e. stems from a mathematical mistake. Furthermore, the evolution kernel (\ref{TrK3}) contains the integral over the spacetime  at fixed $D=4$, while the divergent term has $D \neq 4$. Thus, it cannot even appear within the four-dimensional integral. 

In the dimensional regularization, the power law divergences are discarded ('subtracted'), e.g. \cite{BarVilk-PR1985} and \cite{DeWitt-book2003}, v. 2, p. 547, however, such an ad hoc rule cannot be justified by any physical or mathematical reasons. It is certainly meaningless to separate the four-dimensional logarithmic term from the $D \neq 4$ infinity in (\ref{erreaff}) and then declare it a sought solution. These inconsistencies were corrected in final expressions for the energy-momentum tensor in the covariant perturbation theory \cite{GAV-PLB1993}: the $1/\epsilon$ divergences were discarded  because they are local contributions, and the logarithmic form factors of nonlocal contributions were given the parameter $1/\mu^2=l^2$. Nevertheless, the explained above errors prevented the appearance of the fundamental scaling constant and the two lowest order terms of the covariant action, thus, the gravity theory action, because the first two terms of (\ref{effac3}) were deemed to be nil.

                                                                                    
\end{document}